\documentclass[preprint,prl,aps]{revtex4}
\usepackage{amsmath}
\usepackage{graphicx}
\usepackage{color}
\begin{document}
\title{Tunable asymmetric reflectance in silver films near the percolation threshold}
\author{Aiqing Chen and Miriam Deutsch}
\address{Department of Physics, 1274 University of Oregon, Eugene, OR 97403}

\begin{abstract}
We report on the optical characterization of semicontinuous nanostructured silver films exhibiting tunable optical reflectance asymmetries. The films are obtained using a multi-step process, where a nanocrystalline silver film is first chemically deposited on a glass substrate and then subsequently coated with additional silver via thermal vacuum-deposition. The resulting films exhibit reflectance asymmetries whose dispersions may be tuned both in sign and in magnitude, as well as a universal, tunable spectral crossover point. We obtain a correlation between the optical response and charge transport in these films, with the spectral crossover point indicating the onset of charge percolation. Such broadband, dispersion-tunable asymmetric reflectors may find uses in future light-harvesting systems.
\end{abstract}

\maketitle

\section{I. Introduction}

Nanostructured thin metal films with controllable dispersions are of particular technological interest, due to a number of their highly relevant applications. For example, nanopatterned metal films are known to possess negative refractive indices in the optical domain, potentially allowing imaging with sub-diffraction resolution.\cite{Zhang1,ShalNat} In a different context, studies have shown that the incident-photon-to-current conversion efficiencies of photovoltaic cells with semi-transparent rough metal substrates are significantly higher than those of standard smooth-mirror devices.\cite{Gratzle,Daudrix,Chiba} Enhanced light scattering induced by these so-called hazy substrates leads to multiple scattering, thus enhancing the overall optical path length and absorption efficiency of the trapped light. A significant contribution to the scattering comes from excitation of plasmon resonances in the nanostructured metallic substrates. These resonances are particularly strong when feature size is in the range of several tens of nanometers.\cite{Kreibig}

When designing metal substrates for spectrally-sensitive applications it is important to develop an understanding of the dispersive properties of the composite materials and the role plasmon resonances play in determining the latter. For example, a transmissive hazy substrate is essentially an asymmetric mirror.\cite{Kard} As such, the reflectance of light incident from either side of this mirror is asymmetric, while the transmittance is symmetric. The reflectance $R$ of each side of an asymmetric mirror may vary significantly when the spectrum of the incident radiation is broad enough. Moreover, the \emph{difference}, $\Delta R$ in the two reflectances is typically not constant for a large enough range of wavelengths. We have previously shown that it is possible to fabricate metallic films with dispersion-engineered flat differential reflectance, where the asymmetry of the hazy mirror does not distort the spectrum of the incident white-light radiation.\cite{CHPD} This attribute is highly relevant to solar cell applications, since the asymmetric reflectance as well as the spectral response of the cell window play an important role in determining the overall cell efficiency. One example related to the latter is a recent observation of large asymmetries in photocurrent enhancement mediated by silver nanoparticles positioned either in the rear or the front of Si solar cells.\cite{Catchpole}

Our first proof of principle demonstration of broadband asymmetric reflectance utilized rough nanocrystalline silver films grown by chemical deposition.\cite{JCIS,CHPD} Due to a limited range of control parameters available in this method, metal filling fractions and film thicknesses were the only variables which could be accurately controlled. This led to the emergence of a fixed reflectance asymmetry, whose magnitude and dispersion were invariably determined by the films' microscopic structure. Here we demonstrate how the introduction of an additional design parameter in the form of a thin, vacuum overdeposited silver film allows tuning both the sign as well as the magnitude of the dispersion in the reflectance asymmetry over a wide range of parameters. We present an experimental study of a large number of composite structures comprised of both chemically and vacuum deposited silver films. We show how $\Delta R$ and its dispersion correlate with measured sheet resistances of these films, and demarcate a unique spectral crossover point of the asymmetry as an indication of the onset of charge percolation.

\section{II. Materials and experimental method}

Semi-continuous silver films with controllable filling fractions were deposited on microscope slides using a modified Tollen's reaction as described previously.\cite{CHPD} The amount of silver deposited on the substrates was controlled by monitoring deposition times, with reactions ranging between 1-6 h. Ensuing deposition, coated substrates were rinsed with ultrapure water before being dried with filtered air and stored under nitrogen until tested. Films utilized for optical characterization following this deposition step are referred to here as \emph{single-step} coatings. Some of the chemically deposited films were then coated with an additional film of silver using thermal vacuum deposition at $\approx 10^{-6}$ torr. We refer to these films as \emph{multi-step} coatings. The mass thicknesses of the vacuum deposited overlayers ranged from 10 nm -- 40 nm, as recorded by a calibrated quartz crystal monitor. Following silver deposition samples were stored in inert conditions under nitrogen, to minimize oxidation.

Optical reflectance and transmittance spectra were collected using a spectroscopic optical microscopy setup.\cite{CHPD} A tungsten-halogen white-light source was used to illuminate the samples through an inverted microscope whose output was imaged on the entrance slit of a $F=320$mm spectrometer with a resolution of 0.5 nm. A 10$\times$ (0.25 N.A.) objective was used to collect the normally incident light for spectroscopic imaging onto a liquid nitrogen-cooled detector. A high-reflectance mirror (Newport Broadband SuperMirror, $R\geq99.9\%$) was used to normalize all signals. In addition, data from $\sim1000\mu$m across the films were averaged to obtain each final trace, in order to eliminate spurious effects from local inhomogeneities in the rough films.

\section{III. Results and discussion}

A series of films with varying degrees of metal filling fractions as shown in Fig. 1 were fabricated as described above. Silver filling fractions were measured using digital image analysis of high resolution scanning electron microscope (SEM) micrographs. Images such as in Fig. 1 were first processed using linear contrast enhancement and thresholding. The filling fractions were then measured by pixel counting and linear averaging of the values obtained from several images representative of areas large enough on each sample, such that statistical integrity was maintained. The smallest crystals observed in these films were $\approx5$nm in diameter,\cite{PRL} while the resolution-limited pixel size was found to be $\simeq2$nm. This resulted in an error of only 2--3$\%$ in the final values of the measured filling fractions.

White-light reflection and transmission spectra were acquired for each sample following each silver deposition step. Two separate and averaged reflectance data sets were collected each time, one for light impinging from the metal/air side and the other for light incident from the metal/substrate interface. For consistence we label the former as $R_{1}$ and the latter, acquired after flipping the sample over to its other side, as $R_{2}$. Corresponding transmission spectra, labeled $T_{1}$ and $T_{2}$ were also acquired, using an additional white-light source mounted onto the microscope above the sample, such that the transmitted light was collected by the microscope objective. Here too $T_{1}$ and $T_{2}$ refer to the signal originating from light impinging on the metal/air and metal/substrate interfaces, respectively. As previously, varying degrees of asymmetry were observed in the reflected signals, while the transmittance always remained symmetric, even for highly scattering films with large metal filling fractions.\cite{CHPD}

The reflectance asymmetry is defined as $\Delta R(\lambda)\equiv R_{1}(\lambda)-R_{2}(\lambda)$. Figure 2(a) shows $\Delta R$ plotted against the surface filling fraction, $p$ for ten different single-step samples and measured over the visible spectral range using the spectroscopic microscopy method described above. We find that the asymmetry in $R$, manifest in $\Delta R\neq0$ is weakly dispersive for films with filling fractions $p<0.7$. The dispersion in $\Delta R$, defined as $\partial(\Delta R)/\partial \lambda$ is of fairly constant magnitude in this range of filling fractions. For values of $p$ closer to 1 the dispersion increases dramatically in magnitude while $\Delta R$ attains mostly negative values.  This change is characterized by a broadband spectral crossover point at $p\simeq0.74$, where $\partial(\Delta R)/\partial \lambda=0$ before changing sign.\cite{CHPD} The Inset to Fig. 2(a) shows a typical binary processed image of a single-step film used for determining $p$, the one depicted here having a filling fraction of $p\simeq0.55$. Figure 2(b) shown a typical topography image acquired using an atomic force micrograph (AFM, Digital Instruments Multimode AFM with IIIa controller) of this same sample. Figure 2(c) is a cross-sectional topography analysis, showing the high roughness and discontinuous nature of the chemically deposited film. We find that the rms roughnesses of both single-step and and multi-step films measured by AFM were comparable to their average thicknesses, resulting in very large error in estimating the latter. This is mostly due to the chemical process by which these films are deposited, which renders them rough and highly aggregated. It is known that using measured thicknesses for the characterization or modeling of optical and electronic transport properties of highly granular thin metal films is inaccurate, often resulting in erroneous values of the permittivity~\cite{CHPD} or anomalously high resistivities.\cite{Arnason} In contrast, we have shown that it is possible to utilize the measured surface filling fractions to account for the roughness and porosity of chemically deposited films through geometrical renormalization of their calculated mass thicknesses, without having to directly measure film thickness.\cite{JAP} This renders the metal filling fraction $p$ of single step films an adequate parameter for concise modeling and characterization of the films' properties. We therefore use $p$, which can be consistently measured with much higher accuracy than thickness in these films, as the basis for any further characterization here.

We now address the optical response of our multi-step films. We first examine the dependence of $\Delta R$ on the thickness of the vacuum deposited film in these structures. In Fig. 3(a)-(d) we plot the reflectance asymmetry of four multi-step samples with different metal filling fractions, overcoated with films of various thicknesses, $t$. Data points at zero thickness correspond to single-step coatings before vacuum deposition as in Fig. 2, while the rest depict multi-step films. We divide the data into two distinct groups for analysis - traces taken at wavelengths ranging from 500nm to 800nm, and traces at 400nm and 450nm, where we label the latter the \emph{resonant range}. Examining the first group of traces for $p<0.72$ reveals a behavior of $\Delta R$ similar to that in Fig. 2(a). In each of these figures we identify a crossover point which separates a region of low dispersion in $\Delta R$ from one of opposite sign and higher magnitude. However, the control variable in this case is the mass-thickness of the overdeposited vacuum layers, and not the filling fraction. This leads to the emergence of a \emph{tunable crossover point}, as seen in Fig. 3(a)-(b). At each of these crossover points the dispersion in $\Delta R$ (i.e. $\partial(\Delta R)/\partial\lambda$) is either minimal or even zero. Figure 3(c) depicts the reflectance asymmetry for a sample with $p=0.72$. Close examination reveals a crossover point near zero, at $t\approx2$nm. As we explain below, this is a manifestation of the crossover observed just above $p=0.72$, at $p\simeq0.74$ for the single-step sample in Fig. 2(a). Any addition of overcoat layers only increases the dispersion in $\Delta R$ for that particular filling fraction. For even higher values of $p$ the dispersion in $\Delta R$ increases monotonically with $t$, as shown in Fig. 3(d), and a crossover point does not exist. We discuss the various overcoat thicknesses at which the crossover occurs further below.

Examining the crossover points in Fig. 3 we find them shifting from $t\simeq5$nm in Fig. 3(a) for sparsely filled substrates overcoated with a very thin silver film, through increasing values of overcoat thickness and up to $t\simeq15$nm as in Fig. 3(b) for $p=0.61$. As filling fractions increase above 61$\%$ the crossover point shifts rapidly towards $t=0$nm, vanishing for $p>0.74$. This behavior indicates that metal surface coverage, up to a certain threshold value of $p$ is responsible for the emergence of the observed crossover. Our previous studies have identified a critical value $p_{c}=0.75\pm 0.02$ at which single-step films change from being insulating to conducting electric current. We therefore identify $p_{c}=0.74$ with the percolation threshold in single step, chemically deposited silver films.\cite{Seal,note} To better correlate between the emergence of the crossover and the percolation threshold in our multi-step films we also measured their sheet resistances, first as single step samples as well as after each vacuum deposition of silver. The sheet resistances were obtained using the van der Pauw technique. A Keithley 2400 SourceMeter was used, and the four electrode contacts needed for these measurements were placed at the corners of the substrates, approximately 15 mm apart.\cite{JAP} This large electrode spacing eliminates any finite size effects and shifting of the percolation threshold in these films.\cite{Schmelzer}

Resistance measurements of low filling fraction films with $p=0.15$ and $p=0.20$ reveal that while the single step samples were insulating as anticipated, when overcoated with 10nm of silver their sheet resistances dropped to $26\Omega\pm1\Omega$ and $87\Omega\pm1\Omega$, respectively. This change in resistance is expected, since sparsely coated single step films with $p<p_{c}$ do not have sufficient conducting contacts between silver islands, while the addition of 10nm of silver onto these islands provides the necessary conducting pathways.

Examining films with higher filling fractions (e.g. $p=0.41$ and $p=0.55$) we find that while single step samples are still insulating as expected, deposition of a 10nm overcoating silver film does not result in metallic conductivity. In fact, such samples exhibited low sheet resistances only when the thickness of vacuum deposited layers was 20nm or even 30nm in some cases. This can be explained when we take into account the high roughness that single-step films possess, as can be seen in Fig. 2(c). Due to the large variations in topography and significant fraction of exposed substrate, an overdeposited film with mass-thickness of 10nm may not suffice to form all the conducting pathways necessary for metallic conduction. In such rough films shadowing effects during vacuum evaporation often necessitate deposition of thicker films to achieve charge transport. This also explains the slightly higher sheet resistance measured for the $p=0.20$ film coated with 10nm silver, as compared to a similarly overcoated $p=0.10$ film. As the filling fraction increases from 10\% to 20\% so does film roughness, and consequently also its granular nature.\cite{JAP} While 10nm of thermally deposited silver may suffice to form the necessary conduction pathways in both films, the increasingly high granular nature of films with higher $p$-values results in long and tortuous charge conduction paths, manifest in overall higher measured resistance.\cite{Arnason}

We now address the reflectance asymmetry in the resonant range, as depicted by traces at 440nm and 450nm in Fig. 3(a)-(d). In contrast to the first group discussed above, data traces in the resonant range show consistent departure from the observed trend. This behavior is a signature of the single particle plasmon resonance, situated near $\lambda=450$nm. As expected, the discrepancies are most pronounced in single-step samples with low filling fractions where single particle behavior dominates, as seen in Fig. 3(a). To verify this we plot in Fig. 4(a) the losses, computed as $E_{1,2}=1-R_{1,2}-T_{1,2}$ for each direction of light incidence, for a film with $p=0.15$ and no additional overdeposited silver. We see that both computed traces show a peak near $\lambda=443$nm. This is typical for silver nanoparticles in this size range ($\sim50$nm average particle size, determined from SEM image analysis as discussed above.) For comparison we also plot in Fig. 4(a) the extinction spectrum of an aqueous suspension of silver nanoparticles of similar size fabricated using the same process, measured using a UV/vis spectrometer. The single peak seen in all three traces indicates the excitation of the dipolar plasmon eigenmodes of individual silver nanoparticles. This has also been verified theoretically using Mie scattering formalism.\cite{PRL,Halas} We note that since the measured transmittances $T_{1,2}$ are always symmetric, the discrepancy between $E_{1}$ and $E_{2}$ is a manifestation of the reflectance asymmetry of the structure.

As $p$ increases, particle growth and coalescence typically cause red-shifting and broadening of the single particle plasmon resonance.\cite{Kreibig} The computed losses, $E_{1,2}$ for such a film with $p=0.72$ are shown in Fig. 4(b). However, the increasingly rough nature of our films results in additional intense diffuse scattering of the incident light, which cannot be accounted for by standard Mie scattering theory. Instead, it is common to apply an effective medium theory (EMT) which is used to model the effective linear optical response (e.g the effective real and imaginary components of the permittivity, and the associated effective plasma frequency) of metal films with filling fractions at or above the coalescence regime.\cite{MG,Bruggeman,Abeles,Earp} Albeit, we have demonstrated previously that the commonly used Bruggemann EMT, believed to be most suitable for modeling rough films such as ours, cannot be applied to reliably reproduce the observed losses in these structures, since the microscopic loss mechanisms specific to this model cannot properly reproduce the effect of the high scattering resulting from roughness.\cite{CHPD} In fact, the visible optical response of semicontinuous metal films such as ours, with structural roughness on scales ranging from several nanometers to micrometers cannot be accurately described by \emph{any} EMT which employs a spatially and geometrically averaged response function, replacing the rough film with one having smooth interfaces.\cite{Yagil92} Nevertheless, even if not tractable as here, it is well known that a collective optical response does emerge in metal films near the percolation threshold. In the case of the films studied here this response is manifest in the two data traces in the resonant range tending closer to the first group of longer wavelengths, as seen in Fig. 3(c)-(d). We also see from Fig. 3(a)-(d) that adding overcoat layers has similar impact on the resonant range as increasing $p$ - each additional vacuum deposited film increases the coupling and coalescence of single particles, resulting in an overall collective response and therefore less discrepancies between the two groups of traces at increasing values of $t$.

It is now possible to explain the dependence of the crossover on overcoat thickness as manifestation of the onset of charge percolation. In sparsely coated films the crossover occurs at thinner overcoat thicknesses where the films become conducting, while at intermediate filling ratios (and hence rougher films) slightly thicker overcoat layers are required to achieve onset of charge transport. As the filling fraction increases closer to the percolation threshold, verified experimentally to occur at $p\sim0.74$ in these films, the crossover rapidly shifts towards $t=0$nm as in Fig. 3(c). All samples with $p>0.74$ exhibited sheet resistance values on the order of 10$\Omega$, typical to granular metal films, and accordingly no crossover is observed for these filling ratios at any overcoat thickness.

An important outcome related to the tunable crossover point is the dispersion of the reflectance asymmetry. As discussed previously,\cite{CHPD} the crossover is a manifestation of a broadband (i.e. non-dispersive) asymmetry. In Fig.~5 we plot $\Delta R(\lambda)$ for the same samples as in Fig.~3. While generally monotonic in nature, we find that for samples below the percolation threshold the curves exhibit negative slopes, as in Fig.~5(a) and (b). Traces depicting samples above the percolation threshold cross over to have positive slopes, and hence the opposite sign of dispersion for $\Delta R$. The trace in Fig.~5(c) corresponding to the single-step sample with $p=0.72$ is flat, demonstrating the non-dispersive nature of $\Delta R$ near the crossover point. For comparison we have also plotted $\Delta R(\lambda)$ measured for our vacuum deposited films (Fig.~5(d), Inset.) While the three traces in the Inset corresponding to $t=20$, 30 and 40nm exhibit weak dispersion and are all very close, the data for the thinnest film of $t=10$nm stand out. The latter trace closely resembles the one corresponding to the single-step chemically deposited film in the main Fig.~5(d). Interestingly, the measured resistance of 10nm thick films was comparable in order of magnitude with that of single step chemically deposited films with $p\approx0.80$, ranging from $\sim30-80\Omega$. This is consistent with previous studies showing that very thin, thermally deposited silver films do not exhibit bulk silver behavior, neither in charge transport properties nor optically. We also note that comparison of Fig.~5(a) and the Inset in Fig.~5(d) reveals significant differences in the optical properties of the two systems. While overcoated film thicknesses are equal for the two, the existence of a sparse chemically deposited island-like film significantly perturbs the system at higher overcoat thicknesses. For example, for $t=30$nm we observe that $|\Delta R|$ can be as much as an order of magnitude greater in films containing chemically deposited nanoparticles, even when the particle coverage is sparse as in Fig.~5(a).

\section{IV. Conclusions}

In summary, we demonstrated tuning of the dispersion characteristics of asymmetric mirrors comprised of chemically deposited semicontinuous silver films and coated with additional vacuum-deposited silver. The reflectance asymmetry in such composite coatings was shown to depend on an interplay between initial nanoparticle coverage and overdeposited film thickness, where particle coalescence and coupling play a role in determining the optical response. A tunable spectral crossover point was identified, characterized by vanishing dispersion of the reflectance asymmetry, and it was shown that this crossover indicates the onset of charge transport at the percolation threshold. Dispersion-tunable films fabricated using such multi-step coating techniques may find applications as tunable, spectrally-sensitive substrates and windows in photovoltaic devices, as well as in future engineered metamaterials.

\bigskip

We thank K. Hasegawa for helpful discussions. This work was supported by NSF Grants DMR-02-39273 and DMR-08-04433, and ONAMI ONR.

\newpage

\noindent \textbf{Figure Captions}

\bigskip

\noindent FIG. 1: Scanning electron micrographs of chemically deposited silver films with varying metal filling fractions, as depicted by the value of $p$ in each panel.

\bigskip

\noindent FIG. 2: (a) $\Delta$R as function of filling fraction, plotted across the visible spectrum. Inset: Binary processed image of SEM micrograph of sample with $p\simeq0.55$. (b) Atomic force micrograph of sample with filling fraction $p\simeq0.51$. The vertical separation between the two red arrows corresponds to 0.75$\mu$m. (c) Cross-sectional AFM height analysis of the same sample, the trace between the red arrows corresponding to the segment marked by the arrows in (b). The film's maximal thickness is $\sim$50nm, with an average thickness of $\sim$25nm.

\bigskip

\noindent FIG. 3: $\Delta R$ as function of overcoat thickness $t$ for different filling fractions: (a) $p=0.15$, (b) $p=0.61$, (c) $p=0.72$ and (d) $p=0.80$. The arrows in (a),(b) and(c) indicate the points of minimal dispersion in $\Delta R$, where a crossover occurs.

\bigskip

\noindent FIG. 4: Losses, computed as $E_{1,2}=1-R_{1,2}-T_{1,2}$ for films with (a)$p=0.15$ and (b)$p=0.72$. The dotted traces correspond to losses incurred when the light first impinges on the air/metal interface, and the solid traces depict data for light initially incident from the glass/metal interface. The dashed trace in (a) shows the extinction spectrum of an aqueous suspension of silver nanoparticles of size similar to those in the film and fabricated using the same process, measured using a UV/vis spectrometer.

\bigskip

\noindent FIG. 5: Plots of $\Delta R(\lambda)$ for the same samples as in Fig.~2: (a) $p=0.15$, (b) $p=0.61$, (c) $p=0.72$ and (d) $p=0.80$. The various symbols (circle, square, triangle, star) accompanying the traces in each figure correspond to different overcoat film thicknesses (0, 10, 20 and 30nm, respectively.) Note that in the Inset in (d) showing $\Delta R(\lambda)$ for vacuum deposited films data for a film 40nm in thickness is included.

\newpage
\includegraphics[width=15cm]{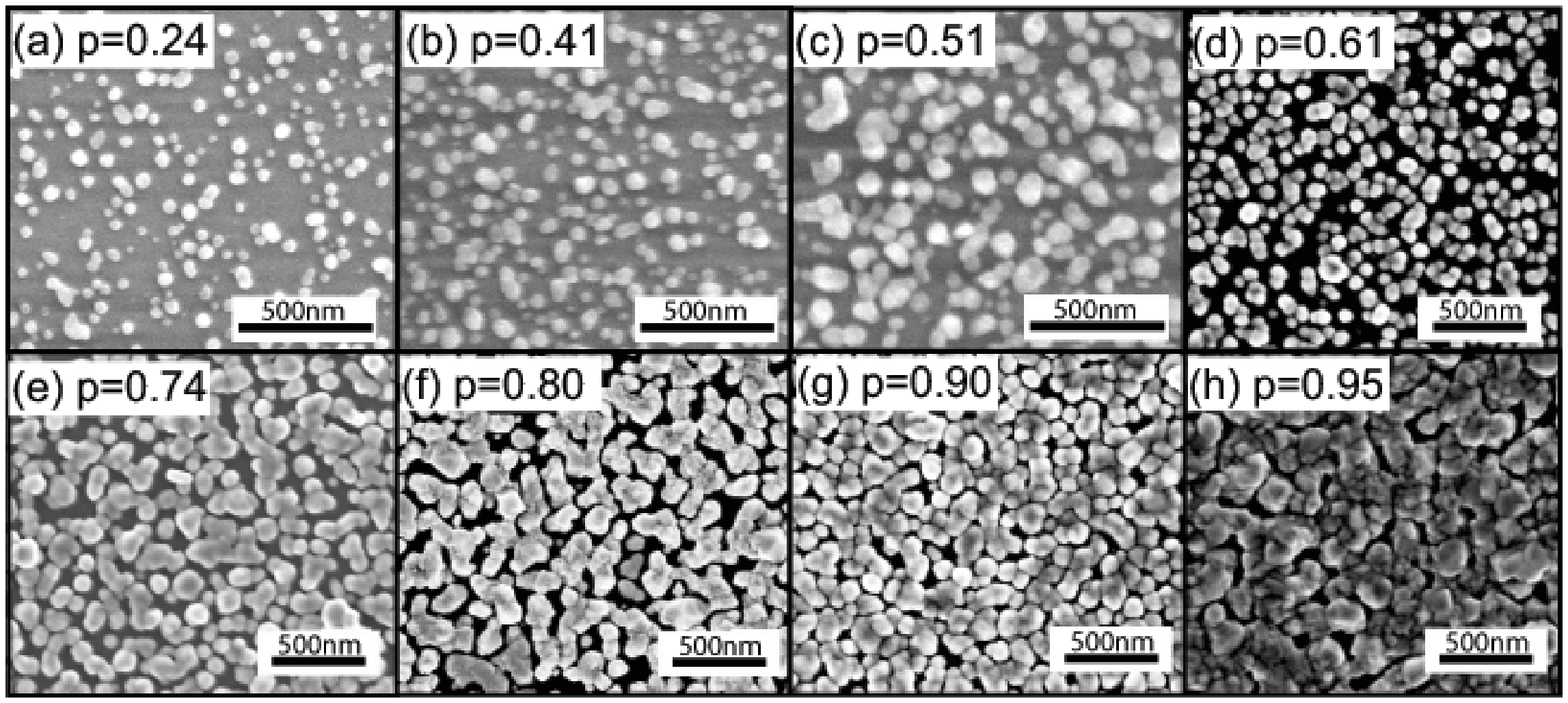}

\newpage
\includegraphics[width=15cm]{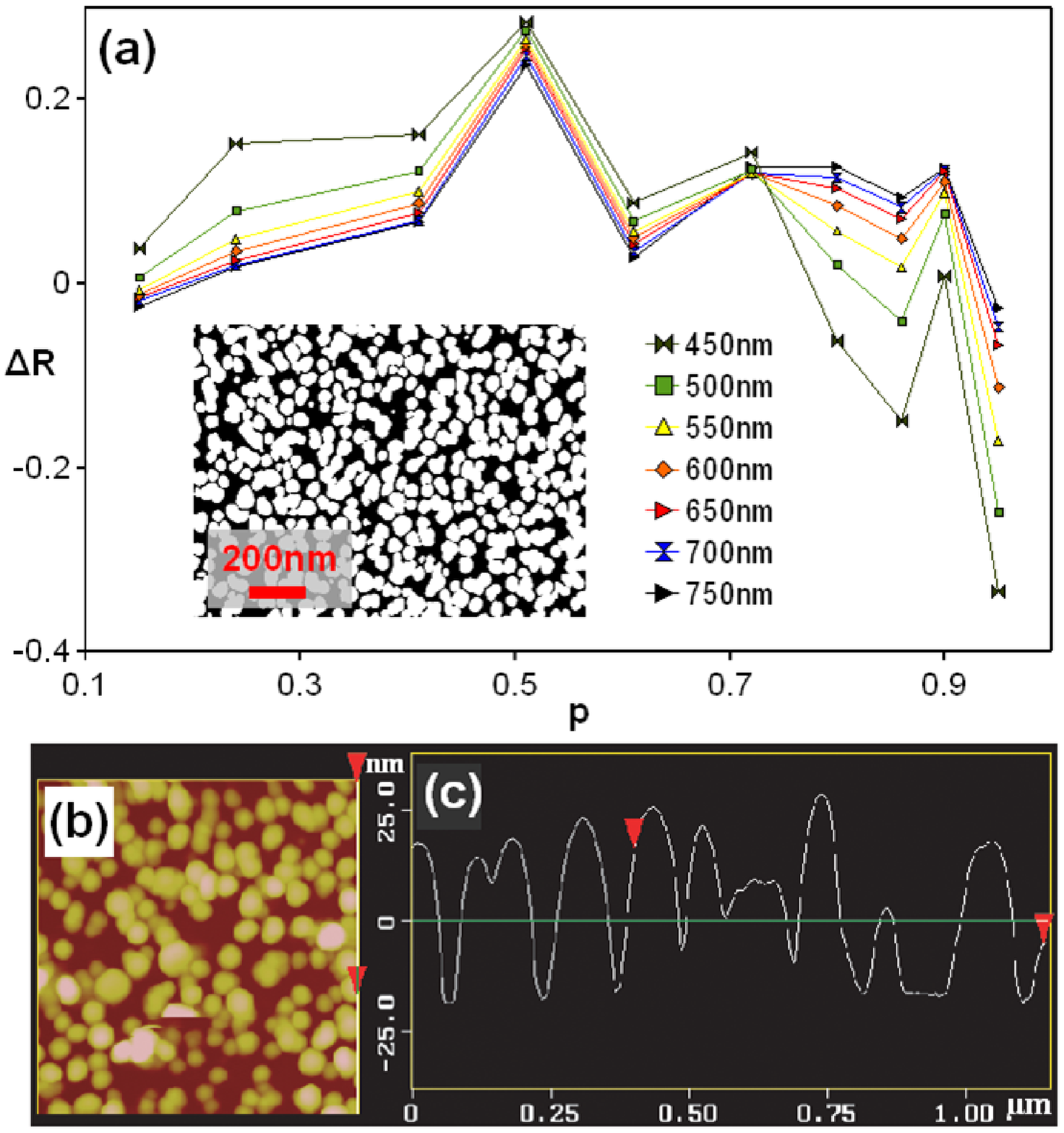}

\newpage
\includegraphics[width=15cm]{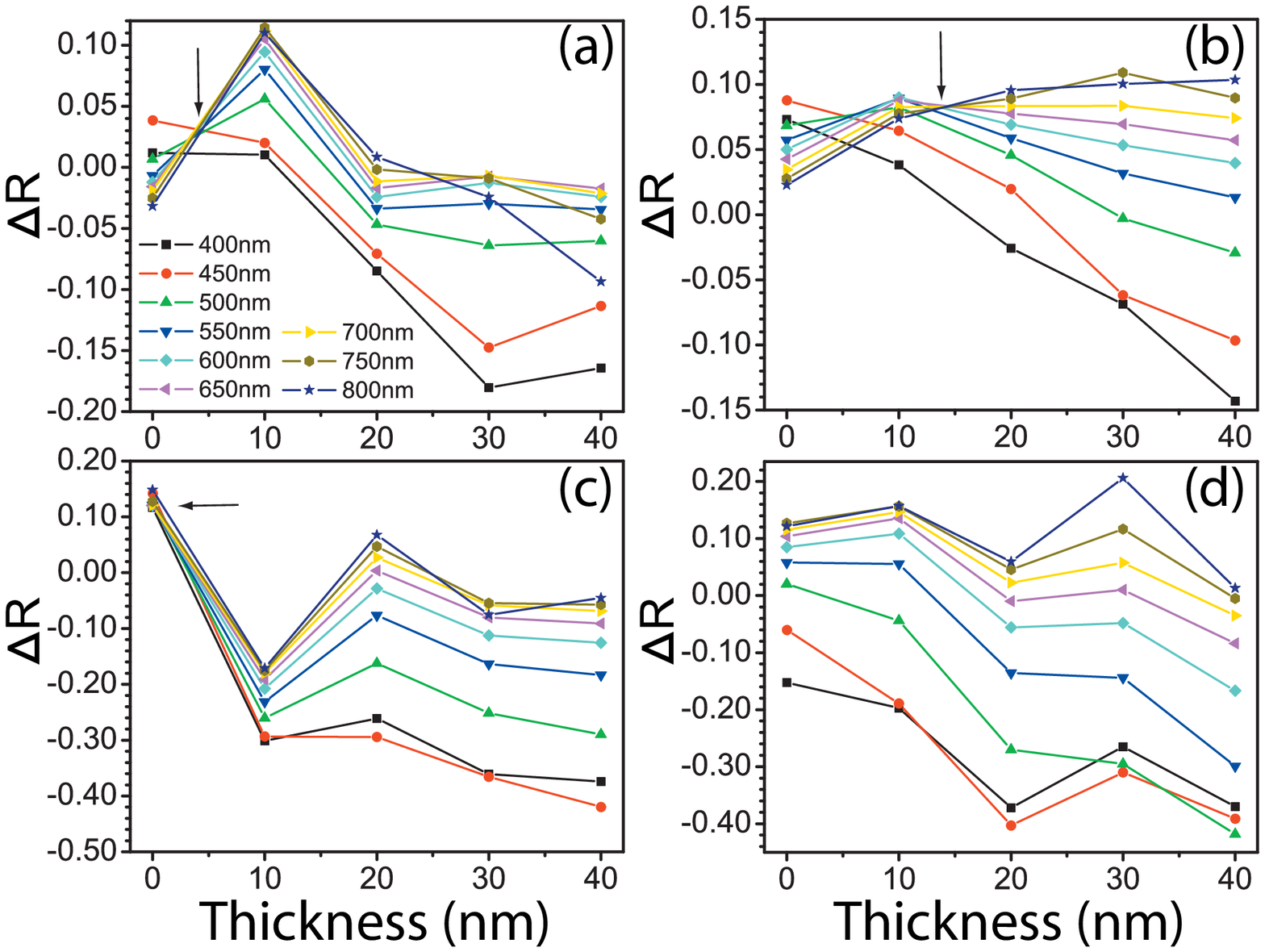}

\newpage
\includegraphics[width=15cm]{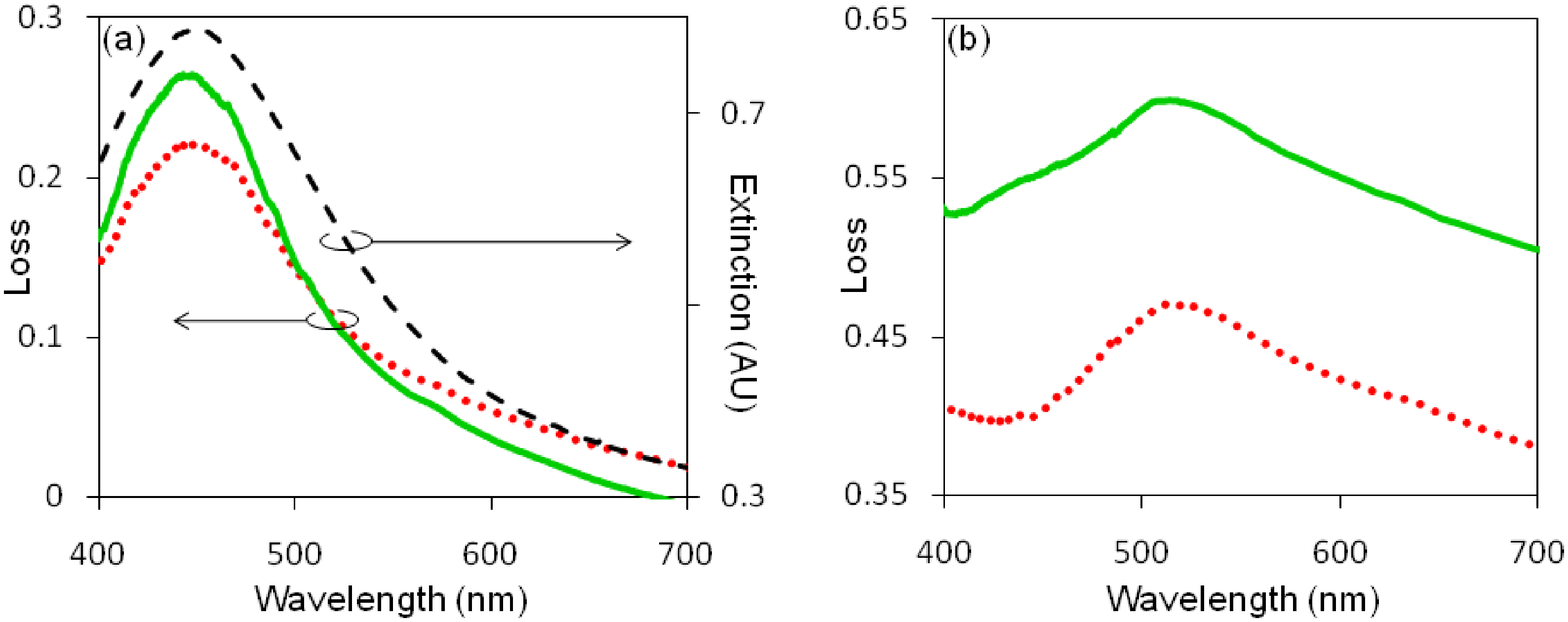}

\newpage
\includegraphics[width=15cm]{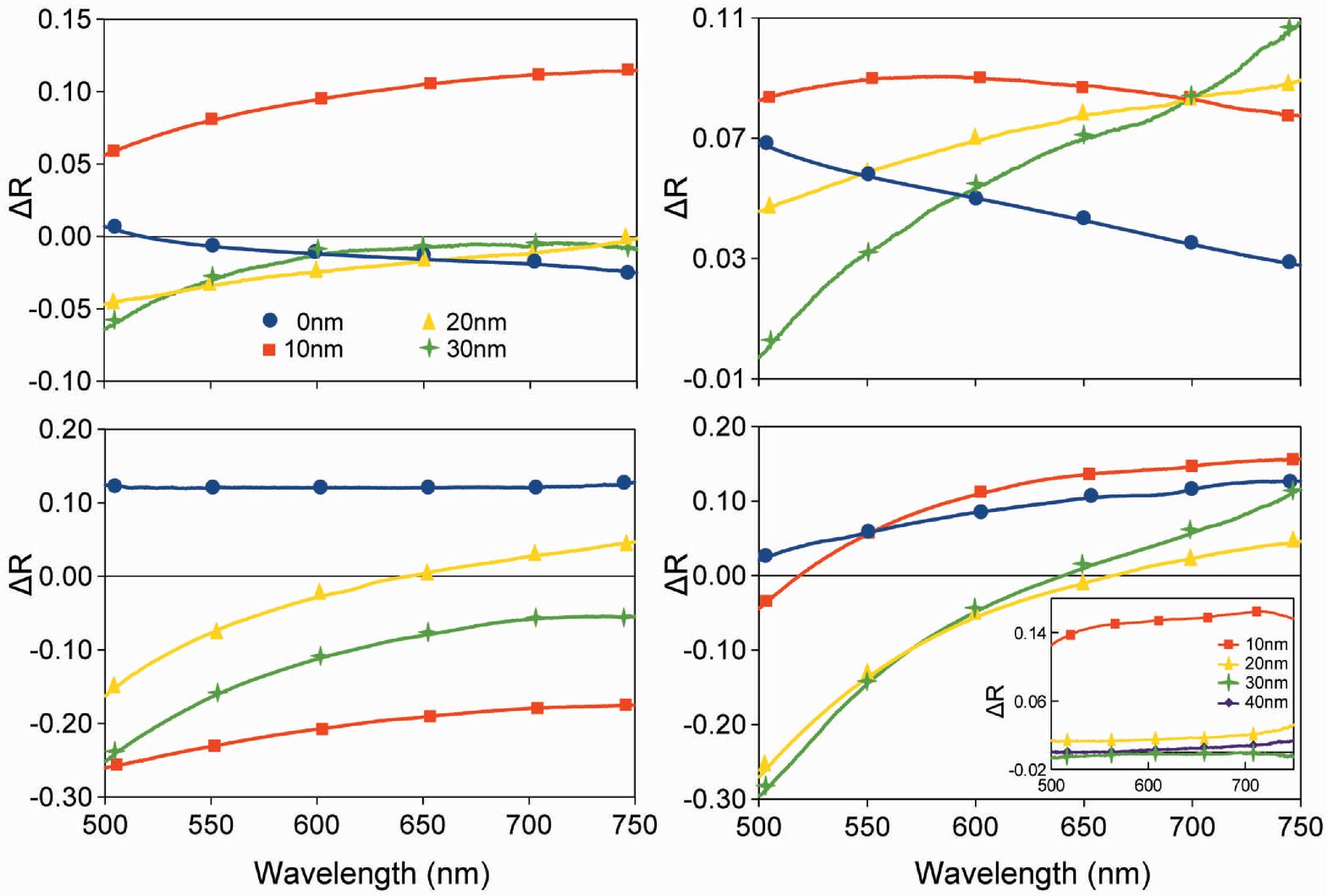}


\begin{thebibliography}{99}
\bibitem{note}
While the filling fraction can be determined with high accuracy from digital image analysis, this method projects the three dimensional structure of the films onto a plane, and therefore tends to overestimate $p$ in denser films.\cite{JAP} It is therefore likely that the actual value of $p$ at the percolation threshold is lower than observed here. In general, it has been previously shown that the percolation threshold is not necessarily unique, and may depend on the particular deposition process.\cite{Seal}
\end{thebibliography}
\end{document}